\documentclass[a4paper,10pt]{article}
\usepackage{amssymb}
\usepackage{eurosym}
\usepackage{amsfonts}
\usepackage{graphicx}
\usepackage{amsmath}

\DeclareFontFamily{U}{rsf}{}
\DeclareFontShape{U}{rsf}{m}{n}{
  <5> <6> rsfs5 <7> <8> <9> rsfs7 <10-> rsfs10}{}
\DeclareMathAlphabet\Scr{U}{rsf}{m}{n} \makeatletter
\@addtoreset{equation}{section} \makeatother

\def\be{\begin{equation}}
\def\ee{\end{equation}}
\def\ba{\begin{array}}
\def\ea{\end{array}}

\newcommand{\bea}{\begin{eqnarray}}
\newcommand{\eea}{\end{eqnarray}}

\begin{document}

\begin{titlepage}
 \thispagestyle{empty}
 \begin{flushright}
     \hfill{DFPD/2016/TH/21}\\
 \end{flushright}

 \vspace{50pt}

 \begin{center}
     { \huge{\bf      {Peccei-Quinn Transformations}}} { \Large{\bf      {and}}} { \huge{\bf      {Black Holes :}}}\\\vspace{5pt} { \huge{\bf      {Orbit Transmutations}}}\\\vspace{10pt}{ \Large{\bf      {and}}} { \huge{\bf      {Entanglement Generation}}}

     \vspace{30pt}

     {\large {\bf Alessio Marrani$^{a,b}$, Thiago Prud\^encio$^{c}$} and {\bf Diego J. Cirilo-Lombardo$^{d,e}$}}

     \vspace{20pt}

  {\it ${}^a$ Museo Storico della Fisica e Centro Studi e Ricerche “Enrico Fermi”,\\
Via Panisperna 89A, I-00184, Roma, Italy.}

\vspace{10pt}

{\it ${}^b$ Dipartimento di Fisica e Astronomia “Galileo Galilei”,
Universit\`a di Padova,\\and INFN, Sez. di Padova,\\Via Marzolo 8, I-35131 Padova, Italy.\\
     \texttt{alessio.marrani@pd.infn.it}}

     \vspace{10pt}

{\it ${}^c$ Coordination of Science \& Technology (CCCT/BICT), Universidade Federal do Maranh\~ao (UFMA)\\
Campus Bacanga, 65080-805, Sao Lu\'is-MA, Brazil.\\
     \texttt{prudencio.thiago@ufma.br}}

     \vspace{10pt}

{\it ${}^d$ National institute of Plasma Physics - INFIP, Consejo Nacional de Investigaciones Cient\'ificas y Tecnicas - CONICET,\\Facultad de Ciencias Exactas y Naturales, Universidad de
Buenos Aires,\\Cuidad universitaria, Buenos Aires 1428, Argentina.\\
     \texttt{diego777jcl@gmail.com}}

      \vspace{10pt}

{\it ${}^e$ Bogoliubov Laboratory of Theoretical Physics, Joint Institute for Nuclear Research\\ 141980,
Dubna-Moscow, Russia.}

     \vspace{30pt}


     {ABSTRACT}
 \end{center}

 \vspace{5pt}
\noindent In a recent paper [Mod. Phys. Lett. \textbf{A30 }(2015) 1550104], the black-hole/qubit correspondence (BHQC) was exploited to define \textquotedblleft black hole quantum circuits" allowing for a change of the supersymmetry-preserving features of electromagnetic charge configurations supporting the black hole solution. This resulted into switching from one $U$-duality orbit to another, or - equivalently - from an element of the corresponding Freudenthal Triple System with a definite rank, to another one. On the supergravity side of BHQC, such quantum gates are related to particular symplectic transformations acting on the black hole charges; namely, such transformations cannot belong to the $U$-duality group, otherwise switching among orbits would be impossible. In this paper, we consider a particular class of such symplectic transformations, namely the ones belonging to the so-called Peccei-Quinn symplectic group, introduced some time ago within the study of very special K\"{a}hler geometries of the vector multiplets' scalar manifolds in $\mathcal{N}=2$ supergravity in $D=4$ spacetime dimensions.
 \vfill
\begin{center}{\sl
Contribution to the \textquotedblleft Universe" special issue titled \textquotedblleft Open Questions in Black Hole Physics"}
\end{center}

\end{titlepage}

\baselineskip 6 mm


\newpage

\newpage

\section{Introduction}\label{Introduction}

In general, extremal black holes (EBHs) appearing as solutions of $D=4$
supergravity theory carry electromagnetic charges that transform linearly
under the $U$-duality\footnote{
Here, $U$-duality is referred to as the ``continuous'' symmetries of \cite{CJ-1,CJ-1-2}.~Their discrete
versions are the $U$-duality non-perturbative string theory symmetries
introduced by Hull and Townsend \cite{HT-1}.}
 group ($G_{4}$), which
however, due to its non-transitive action, determines a stratification of
the corresponding representation space $\mathbf{R}$ \cite{FG} (cf.,~e.g., see \cite{Small-Orbits, Marrani-review} for reviews and lists of references).
Due to the attractor mechanism \cite{AM-Refs,AM-Refs-2,AM-Refs-3,AM-Refs-4,AM-Refs-5}, the~Bekenstein--Hawking \cite{BH,BH-2} entropy of an EBH is a function only of the electromagnetic charges; in
most of the $\mathcal{N}=2$ models with symmetric scalar manifolds, as well as in
$\mathcal{N}>3$-extended theories, the~entropy is given by ($\pi $ times)
the square root of the absolute value of the $U$-duality quartic invariant $%
\mathcal{I}_{4}$ \cite{Kallosh-Kol, Ferrara-Gimon-Kallosh} (also see, {%
e.g.,} \cite{Kallosh-review, Hayakawa} for introductory reviews and lists of
references).

Due to the existence of the Bogomol'nyi--Prasad--Sommerfield (BPS) bound, the
mass of an EBH is bounded from below; for instance, in the Reissner--Nordstr%
\"{o}m BH, which can be embedded in $\mathcal{N}=2$ ``pure''
supergravity, as well as in $\mathcal{N}=1$ supergravity coupled to one
vector multiplet~\cite{N=1-Attractors}, $M\geq \sqrt{q^{2}+p^{2}}$ (which is
saturated in the extremal case: $M=\sqrt{q^{2}+p^{2}}$). On the other hand,
in~the presence of scalars, extremality does not necessarily imply the saturation
of the BPS bound: the simplest example is provided by the dilaton EBH
(which has a~natural embedding into $\mathcal{N}=4$ ``pure'' supergravity),
for which $M\geq \frac{1}{\sqrt{2}}(|Q|+|P|)$ generally holds, with~$Q$ and $%
P$ denoting suitably dilaton-dressed electric and magnetic charges \cite%
{kallosh1} (see also \cite{Hayakawa}); in this case, the~Bekenstein--Hawking
entropy reads $S=2\pi |PQ|$.

An EBH can be conceived as the final limit state of a thermal evaporation
process associated with the emission of Hawking radiation; when reaching the
extremal state (characterized by vanishing temperature), the Hawking
radiation ceases: the EBH is not evaporating anymore, in order to sustain
the consistency of the BPS bound \cite{kallosh1}. If the BH charge is large,
the last stages of evaporation occur for a mass much greater than Planck
mass $M_{P}$, but the EBH can then discharge by the creation of pairs of
elementary charged particles \cite{wilczek,wilczek2}; for example, in the
case of a dyonic Reissner--Nordstr\"{o}m BH, the~evaporation stops when the
charge achieves the absolute value of Planck mass: $\sqrt{q^{2}+p^{2}}%
=M_{P} $.

In the four-dimensional Einstein gravity theory with maximal local
supersymmetry, namely $\mathcal{N}=8$ supergravity, various degrees of BPS
supersymmetry preservation are possible, each one associated with a~massive
representation of the $\mathcal{N}=8$ supersymmetry algebra \cite{ferrara2,
Nahm}: namely, $1/2$-BPS, $1/4$-BPS and $1/8$-BPS EBHs exist~\cite%
{Ferrara-Maldacena, FG} (cf. also \cite{N=8-Observations} for a
more recent analysis). Essentially, the~classification of BPS states
preserving different amounts of supersymmetry of the background algebras is
similar to the stratification of timelike, light-like and spacelike vectors
in a Minkowski space \cite{FG}; in such a classification, EBHs are usually
named ``small'' or ``large'', depending
on whether their Bekenstein--Hawking entropy vanishes or not.

The string-theoretic interpretation of BHs is usually given in terms of $Dp$%
-branes; their wrapping around compactified dimensions allows for a
consistent picture, relating them to qubits (the basic quantities studied by
quantum information theory (QIT)): this led to the formulation of the
so-called {black-hole/qubit correspondence (BHQC}){\cite{Duff, FD-BHQC-1,borsten1,borsten0,FD-BHQC-2,borsten11, borsten12,levay4,borsten4,levay3,levay5,
levay6}}. 
Important achievements have been reached in this context, and~some
others are underway; just to name a few, the EBH entropy in the so-called $%
\mathcal{N}=2$ supergravity STU
 model \cite{STU,STU-2} was associated with
tripartite entanglement measurement and classification in QIT \cite%
{kallosh,Duff}, and~this was further extended to the hitherto unsolved issue
of the classification of four qubits entanglement \cite{borsten13}.
Moreover, the~Hilbert space for qubits was traced back to wrapped branes
inside the cohomology of the extra dimensions~\cite{levay,borsten12}, and
exceptional groups $E_{7}$ and $E_{6}$ were respectively related to the
tripartite entanglement of seven qubits \cite{FD-E7} and to the bipartite
entanglement of three qutrits \cite{FD-E6}. Recently, ``BH
quantum gates'' were formulated within the BHQC in \cite{prudencio1}.

The Freudenthal triple system (FTS) \cite{FTS,FTS-2,FTS-3} plays a key role in BHQC; these are
algebraic systems, whose~elements can be grouped into different ranks,
ranging from zero to four \cite{ferrar, Krutelevich,Krutelevich-2}. In the classification of
tripartite entanglement, the ranks $0$, $1$, $2a$, $2b$, $2c$, $3$ and $4$
are associated respectively with null, separable $a-b-c$, biseparable $a-bc$, $%
b-ac$ and $c-ab$, W and GHZ
 states \cite{borsten0}. By means of a set of
local operations and classical communications (LOCC), the state can be
operated without generating a~new class of entanglement. The inter-relation
between equivalent states in the same class is achieved by the possibility
of interconverting one to another by means of the group of {local
unitaries} (LU) in the set of LOCC \cite{local unitaries}. Rather than
declare equivalence when states are deterministically related to each other
by LOCC, one may require only that they may be transformed into one another
with some non-zero probability of success; this can be achieved by the
so-called stochastic LOCC (SLOCC)~\cite{SLOCC}. Within the SLOCC paradigm,
two states are identified if there is a non-zero probability that one can be
converted into the other and vice versa; one can trace back the origin of
such a generalization by the physically-motivated fact that any set of SLOCC
equivalent states may be used to perform the same non-classical operations,
only with varying likelihoods of success. In terms of FTS, with SLOCC, one is
varying the representative element with the same entanglement class,
corresponding to a well-defined FTS rank. As~analyzed in \cite{borsten1},
SLOCC equivalence classes of three qubit states are stratified into an~entanglement hierarchy~\cite{SLOCC}, which can be consistently reproduced in
terms of FTS, or equivalently in terms of orbits of its automorphism~group.

Going {beyond} the SLOCC paradigm within the BHQC, \cite{prudencio1}
investigated the possibility of establishing quantum circuits (realizing
certain classes of quantum gate operations) capable of switching, within the
same FTS, from one rank class to another one. On the BH (supergravity) side
of BHQC, this essentially amounts to switching from one $U$-duality
orbit to
another one; in the present treatment, this phenomenon will be named {%
orbit} {transmutation}. In FTS terms, this amounts to switching from
an FTS element with a definite rank to another FTS element with another
rank, lying in another orbit of the FTS automorphism group. In this context,
specific quantum circuits generating Bell states, quantum teleportation and
GHZ states can be clearly associated with the protocols of actions on the charge
state configurations of EBHs \cite{prudencio1}.

On the BH (supergravity) side of BHQC, the aforementioned ``BH quantum gates'' are related to a~particular kind of symplectic
transformations (which are the most general transformations acting linearly
on BH electromagnetic charges \footnote{
symplectic transformations also provide an example of pseudo-dualities in
supergravity~\cite{Hull-VP}}).
 Indeed, ``BH quantum gate''
symplectic transformations generally allow for the switching from the
representative of one $U$-duality orbit to a representative of {%
another} $U$-duality orbit; the most interesting cases are the ones in which
a change of FTS rank occurs, {i.e.,} in which a transition between
{different} levels of the entanglement hierarchy takes place. Since
the $U$-duality orbit structure (namely, the stratification of the charge
representation space $\mathbf{R}$ under the non-transitive action of the $U$%
-duality group $G_{4}$) is invariant under $G_{4}$ itself, it is clear that
such symplectic transformations cannot belong to the $U$-duality group,
otherwise switching among orbits would be impossible. Thus, the
``BH quantum gate'' transformations generally belong to the
pseudo-Riemannian, non-symmetric coset \footnote{
this coset was recently exploited in the analysis of the so-called
symplectic deformations of gauged $\mathcal{N}=8$, $D=4$ supergravity \cite%
{Sympl-Def}, later extended to other supergravity~theories}.
\begin{equation}
\frac{Sp(2n+2,\mathbb{R})}{G_{4}}, \label{Sp/G}
\end{equation}%


where
$2n+2$ denotes the dimension of the charge representation space $\mathbf{R}$.\medskip

In the present paper, we consider a particular class of ``BH
quantum gate'' symplectic transformations, namely the ones belonging to the
so-called {Peccei--Quinn (PQ) symplectic group}, introduced in {\cite{
PQ}}
 in the study of very special ({i.e.}, cubic) K\"{a}hler
geometries in $\mathcal{N}=2$ supergravity \footnote{
We will always consider the ''large, real charge''
supergravity limit within BHQC. In the case of (dyonic) quantized charges,
the analysis of FTSs is more complicated, and a full classification of $U$%
-duality orbits is not even currently available (for some advances along
this venue, and lists of references, cf., e.g., \cite{FD-Duff,
N=8-Observations}).}. In such a framework, the special K\"{a}hler geometry
of the vector multiplets' scalar manifold is based on a holomorphic
prepotential function $F$ in the symplectic sections $X^{\Lambda }$. {%
At~least} within models with symmetric scalar manifolds, all symplectic
transformations belonging to the coset Equation (\ref{Sp/G}) and, thus, {a
fortiori} the PQ symplectic transformations, change the value of unique
quartic invariant polynomial $\mathcal{I}_{4}$ of the charge representation
space $\mathbf{R}$, and consequently, they change the EBH Bekenstein--Hawking
entropy. This is of particular importance in the theory of EBH attractors,
because such a phenomenon allows for {orbit transmutations}, namely
for transitions between different charge orbits in the stratification of $%
\mathbf{R}$, with a corresponding change of the supersymmetry-preserving
features of the corresponding EBH charge configurations \cite{PQ}.

The plan of the paper is as follows.

In Section \ref{Sec-2}, we recall the definition of the Peccei--Quinn symplectic
group and introduce the corresponding operator in the Hilbert space of EBH
charge configuration states. Then, in Section \ref{Sec-3}, we introduce some
examples of ``large'' and ``small'' EBH charge
configurations, whose transformation under the PQ operator is discussed in
Section \ref{Sec-4}, with particular emphasis on the occurrence of different
types of orbit transmutations. After a brief interlude dedicated to state
superposition in Section \ref{Sec-5}, in Section \ref{Sec-6}, we start the analysis
of general mechanisms of entanglement generation on EBH charge states, by
defining the entanglement PQ operator and pointing out the necessity of
switching to a~complex ground field (a already stressed within BHQC; see,
 {e.g.,} \cite{borsten0}) within the indistinguishability assumption.
Some final comments are contained in Section \ref{Conclusion}.

\section{Peccei--Quinn Symplectic Group and Operator} \label{Sec-2}

The transformations of the Peccei--Quinn (PQ) group $PQ\left( n+1\right) $
were introduced in \cite{PQ}, \mbox{as given by:}
\begin{equation}
\mathbf{W:=}\left(
\begin{array}{cc}
\mathbb{I}_{n+1} & 0_{n+1} \\
\mathcal{W} & \mathbb{I}_{n+1}%
\end{array}%
\right) \in PQ\left( n+1\right) , \label{W}
\end{equation}%
where $\mathbb{I}_{n+1}$ and $0_{n+1}$ respectively denote the identity and
null $(n+1)\times (n+1)$ matrices and $\mathcal{W}$ is defined as the
following $(n+1)\times (n+1)$ real matrix ($i,j=1,...,n$, $\Theta
_{ij}=\Theta _{\left( ij\right) }$):
\begin{equation}
\mathcal{W}:=\left(
\begin{array}{cc}
\varrho & c_{j} \\
c_{i} & \Theta _{ij}%
\end{array}%
\right). \label{W-call}
\end{equation}%

It is then easy to realize that $PQ\left( n+1\right) $ is an $(n+1)(n+2)/2$%
-dimensional Abelian group given by:
\begin{equation}
PQ\left( n+1\right) =Sp(2n+2,\mathbb{R})\cap LUT(2n+2,\mathbb{R}),
\end{equation}%
where $Sp(2n+2,\mathbb{R})$ is the maximally non-compact (split) real form
of the symplectic group whose Lie algebra is $\mathfrak{c}_{n+1}$ (in Cartan
notation), and $LUT(2n+2,\mathbb{R})$ denotes the subgroup of $SL(2n+2,%
\mathbb{R})$ given by the $\left( 2n+2\right) \times \left( 2n+2\right) $
lower unitriangular matrices, which are unipotent.

The group $PQ\left( n+1\right) $ and its Lie algebra $\mathfrak{pq}_{n+1}$
were introduced in \cite{PQ} in the study of very special K\"{a}hler
geometries. However, interestingly, matrices with structure as $\mathbf{W}$ (%
\ref{W}), and thus belonging to $PQ\left( n+1\right) $, made their
appearance also in other contexts. For example, $\mathbf{W}$ can be regarded
as a~particular case of the quantum perturbative duality transformations in
supersymmetric Yang--Mills theories coupled to supergravity \cite{41}; in~such a framework, Equation (\ref{W}) defines the structure of quantum
perturbative monodromy matrices in heterotic string compactifications with
classical $U$-duality group $SL(2,\mathbb{R})\times SO(2,n+2)$.

In $\mathcal{N}=2$ (ungauged) supergravity theories in $D=4$ space-time
dimensions, $n$ denotes the number of (Abelian) vector multiplets coupled to
the gravity multiplet. By virtue of a theorem due to Dynkin~\cite{Dynkin},
the $U$-duality group $G_{4}$ is embedded into the symplectic group $Sp(2n+2,%
\mathbb{R})$ in a~maximal (and generally non-symmetric) way, as follows:
\begin{eqnarray}
G_{4} &\subsetneq &Sp(2n+2,\mathbb{R}); \\
\mathbf{R} &=&\mathbf{2n+2},
\end{eqnarray}%
where $\mathbf{R}$ is the symplectic ({i.e.}, anti-self-conjugated)
representation of $G_{4}$ in which the electric and magnetic charges of the
(E)BH sit. Consequently, one can define a dyonic\footnote{
We use the so-called {''special coordinates''}
symplectic frame of $\mathcal{N}=2$, $D=4$ supergravity \cite{central
extension}. The PQ symplectic group was actually introduced within the
so-called $4D/5D$ {``special coordinates''}
symplectic frame, which is the most natural for the study of (projective)
very special K\"{a}hler geometry \cite{PQ}.} charge vector state $|\mathcal{Q%
}\rangle $ as follows\footnote{
Throughout the present investigation, we will not make use of the Einstein
summation convention. Such a choice, which may result in being cumbersome for the
customary supergravity treatment, is made in order to comply with the most
used notation in QIT.}:
\begin{equation}
|\mathcal{Q}\rangle:=p^{0}|0\rangle +q_{0}|1\rangle
+\sum_{i=1}^{n}p^{i}|2^{i}\rangle +\sum_{i=1}^{n}q_{i}|3^{i}\rangle ,
\label{Q-state}
\end{equation}%
where:
\begin{equation}
\{|0\rangle ,|1\rangle ,|2^{i}\rangle ,|3^{i}\rangle ,i=1,...,n\}
\end{equation}%
is a symplectic basis of the Hilbert space realization of $\mathbf{R}$.
It is instructive to report here the holomorphic sections' state associated
with a given holomorphic prepotential $F\left( X\right) $:
\begin{equation}
|\mathcal{H}\rangle =X^{0}|0\rangle +F_{0}|1\rangle
+\sum_{i=1}^{n}X^{i}|2^{i}\rangle +\sum_{i=1}^{n}F_{i}|3^{i}\rangle ,
\label{H-state}
\end{equation}%
where $F_{i}(X):=\partial F(X)/\partial X^{i}$.

A general operator in the Hilbert space of EBH state configurations can be
defined as follows:
\begin{eqnarray}
\hat{\mathcal{O}}_{BH} &:&=\sum_{\alpha ,\beta =0}^{1}s_{\alpha \beta
}|m\rangle \langle n|+\sum_{i=1}^{n}\sum_{\alpha =0}^{1}\left( r_{i\alpha
}|2^{i}\rangle \langle \alpha |+t_{i\alpha }|3^{i}\rangle \langle \alpha
|+r_{\alpha i}|\alpha \rangle \langle 2^{i}|+t_{\alpha i}|\alpha \rangle
\langle 3^{i}|\right) \notag \\
&+&\sum_{i,j=1}^{n}\left( a_{ij}|2^{j}\rangle \langle
2^{i}|+b_{ij}|3^{j}\rangle \langle 3^{i}|+c_{ij}|2^{j}\rangle \langle
3^{i}|+d_{ij}|3^{j}\rangle \langle 2^{i}|\right). \label{O-BH}
\end{eqnarray}%

However, in the present treatment, we will focus on the class of {%
Peccei--Quinn (PQ) operators}, related to the aforementioned PQ symplectic
transformations. From Equations (\ref{W}) and (\ref{W-call}), one can define the PQ
operator $\mathcal{O}_{PQ}$ associated with the PQ transformation $\mathbf{W}$
as follows:
\begin{equation}
\hat{\mathcal{O}}_{PQ}:=|0\rangle \langle 0|+|1\rangle \langle
1|+\sum_{i=1}^{n}\left( |2^{i}\rangle \langle 2^{i}|+|3^{i}\rangle \langle
3^{i}|\right) +\rho |1\rangle \langle 0|+\sum_{i=1}^{n}\left( c_{i}|1\rangle
\langle 2^{i}|+c_{i}|3^{i}\rangle \langle 0|\right) +\sum_{i,j=1}^{n}\Theta
_{ij}|3^{i}\rangle \langle 2^{j}|, \label{O-PQ}
\end{equation}%
corresponding to:%
\begin{equation}
\forall i,j=1,..,n:\left\{
\begin{array}{l}
s_{00}=1=s_{11},~~s_{10}=\rho ,~~s_{01}=0; \\
r_{i0}=0=r_{i1},~~t_{i0}=c_{i},~~t_{i1}=0; \\
r_{0i}=c_{i},~~r_{1i}=0,~~t_{0i}=0=t_{1i}; \\
a_{ij}=\delta _{ij}=b_{ij},~~c_{ij}=0,~~d_{ij}=\Theta _{ij}=\Theta _{(ij)}%
\end{array}%
\right.
\end{equation}%
in the parametrization of the general operator (\ref{O-BH}).

The PQ-transformations of States (\ref{Q-state}) and (\ref{H-state})
formally read as:
\begin{eqnarray}
|\mathcal{Q}_{PQ}\rangle &:&=\hat{\mathcal{O}}_{PQ}|\mathcal{Q}\rangle ;
\label{Q-PQ} \\
|\mathcal{H}_{PQ}\rangle &:&=\hat{\mathcal{O}}_{PQ}|\mathcal{H}\rangle.
\end{eqnarray}

\section{Some ``Large'' and ``Small'' Configurations}\label{Sec-3}

Within the considered symplectic frame, the general charge configuration
state (\ref{Q-state}) is associated with the following, unique \cite%
{Sato-Kimura, Kac} $G_{4}$-invariant ({cf., e.g.,} \cite{FG,
Ferrara-Gimon-Kallosh, Special-Road} and the refrences therein):%
\begin{equation}
\mathcal{I}_{4}(\mathcal{Q})=-\left(
p^{0}q_{0}+\sum_{i=1}^{n}p^{i}q_{i}\right) ^{2}+4q_{0}\mathcal{I}%
_{3}(p)-4p^{0}\mathcal{I}_{3}(q)+4\{\mathcal{I}_{3}(p),\mathcal{I}_{3}(q)\},
\end{equation}%
where:
\begin{eqnarray}
\mathcal{I}_{3}(p) &:&=\frac{1}{3!}%
\sum_{i,j,k=1}^{n}d_{ijk}p^{i}p^{j}p^{k},~~~~\mathcal{I}_{3}(q):=\frac{1}{3!}%
\sum_{i,j,k=1}^{n}d^{ijk}q_{i}q_{j}q_{k}, \\
\{\mathcal{I}_{3}(p),\mathcal{I}_{3}(q)\} &:&=\sum_{i=1}^{n}\frac{\partial
\mathcal{I}_{3}(p)}{\partial p^{i}}\frac{\partial \mathcal{I}_{3}(q)}{%
\partial q_{i}}=\frac{1}{4}%
\sum_{i,j,k,l,m=1}^{n}d^{ijk}d_{ilm}q_{j}q_{k}p^{l}p^{m}.
\end{eqnarray}%

As anticipated in Section \ref{Introduction}, the Bekenstein--Hawking EBH
entropy formula can be recast in terms of such an~invariant as:
\begin{equation}
S(\mathcal{Q})=\pi \sqrt{|\mathcal{I}_{4}(\mathcal{Q})|}.
\end{equation}

Below, we consider some noteworthy EBH charge configurations, which also
provide well-defined {representatives of the corresponding $U$-duality
orbits, {i.e.}, examples of FTS elements of all possible ranks~(\mbox{$4$, $%
3 $, $2$, $1$}).}

\begin{enumerate}
\item \textbf{Kaluza-Klein (KK) configurations}:
\begin{eqnarray}
|\mathcal{Q}_{KK}\rangle &:&=p^{0}|0\rangle +q_{0}|1\rangle ; \\
\mathcal{I}_{4}(\mathcal{Q}_{KK}) &=&-\left( p^{0}\right) ^{2}q_{0}^{2}<0, \\
\frac{\partial \mathcal{I}_{3}(q)}{\partial q_{i}} &=&0,\frac{\partial
\mathcal{I}_{3}(p)}{\partial p^{i}}=0,~\forall i, \\
S(\mathcal{Q}_{KK}) &=&\pi |p^{0}q_{0}|.
\end{eqnarray}%
\textbf{1.1]} 
 In the case with both $p^{0}$ and $q_{0}$ non-vanishing, one
obtains a dyonic ``large'' configuration (corresponding to
the maximal rank$=4$ element in the related FTS), whereas for $p^{0}$ or $q_{0}$
vanishing, one has a ``small'' configuration (namely: \textbf{%
1.2]} for $p^{0}=0$: $|\mathcal{Q}_{eKK}\rangle $ and \textbf{1.3]} for $%
q_{0}=0$: $|\mathcal{Q}_{mKK}\rangle $, such~that $S(\mathcal{Q}_{eKK})=S(%
\mathcal{Q}_{mKK})=0$, corresponding to minimal rank$=1$ elements in the
FTS).

\item \textbf{Electric (E) configurations }:%
\begin{eqnarray}
|\mathcal{Q}_{E}\rangle &:&=p^{0}|0\rangle +\sum_{i=1}^{n}q_{i}|3^{i}\rangle
; \\
\mathcal{I}_{4}(\mathcal{Q}_{E}) &=&-4p^{0}\mathcal{I}_{3}(q)\neq 0, \\
\frac{\partial \mathcal{I}_{3}(p)}{\partial p^{i}} &=&0, \\
S(\mathcal{Q}_{E}) &=&2\pi \sqrt{\left\vert p^{0}\mathcal{I}%
_{3}(q)\right\vert }.
\end{eqnarray}%
\textbf{2.1]} In the case with both $p^{0}$ and $\mathcal{I}_{3}(q)$
non-vanishing, one obtains a dyonic ``large'' configuration
(corresponding to the maximal rank$=4$ element in the related FTS, {i.e.}%
, to an element in the same duality orbit of the ``large''
dyonic KK configuration). As an example of ``small''
configurations ($S(\mathcal{Q}_{E})=0$), one~may set $p^{0}=0$; this latter
case further splits into three subcases: \textbf{2.2]} $\mathcal{I}%
_{3}(q)\neq 0$, corresponding to a~rank-three FTS element: $|\mathcal{Q}%
_{3E}\rangle $; \textbf{2.3]} $\mathcal{I}_{3}(q)=0$, but $\partial \mathcal{%
I}_{3}(q)/\partial q_{i}\neq 0$ for {at least} some $i$,
corresponding to a rank-two FTS element: $|\mathcal{Q}_{2E}\rangle $;
\textbf{2.4]} $\partial \mathcal{I}_{3}(q)/\partial q_{i}=0~\forall i$, but $%
q_{i}\neq 0$ for {at least} some $i$, corresponding to a rank-one FTS
element ({i.e.},~in~the same duality orbit of the small KK
configurations): $|\mathcal{Q}_{1E}\rangle $.

\item \textbf{Magnetic (M) configurations} :%
\begin{eqnarray}
|\mathcal{Q}_{M}\rangle &=&q_{0}|1\rangle +\sum_{i=1}^{n}p^{i}|2^{i}\rangle ;
\\
\mathcal{I}_{4}(\mathcal{Q}_{M}) &=&4q_{0}\mathcal{I}_{3}(p)\neq 0, \\
\frac{\partial \mathcal{I}_{3}(q)}{\partial q_{i}} &=&0, \\
S(\mathcal{Q}_{M}) &=&2\pi \sqrt{\left\vert q_{0}\mathcal{I}%
_{3}(p)\right\vert }.
\end{eqnarray}%
\textbf{3.1]} In the case with both $q_{0}$ and $\mathcal{I}_{3}(p)$
non-vanishing, one obtains a dyonic `` large'' configuration
(corresponding to the maximal rank$=4$ element in the related FTS). As an
example of ``small'' configurations ($S(\mathcal{Q}_{M})=0$),
one may set $q_{0}=0$; this latter case further splits into three subcases:
\textbf{3.2]} $\mathcal{I}_{3}(p)\neq 0$, corresponding to a rank-three FTS
element: $|\mathcal{Q}_{3M}\rangle $; \textbf{3.3]} $\mathcal{I}_{3}(p)=0$,
but $\partial \mathcal{I}_{3}(p)/\partial p^{i}\neq 0$ for {at least}
some $i$, corresponding to a rank-two FTS element: $|\mathcal{Q}%
_{2M}\rangle $; \textbf{3.4]} $\partial \mathcal{I}_{3}(p)/\partial
p^{i}=0~\forall i$, but~$p^{i}\neq 0$ for {at least} some $i$,
corresponding to a rank-one FTS element: $|\mathcal{Q}_{1M}\rangle $.
\end{enumerate}

\section{Peccei--Quinn Orbit Transmutations} \label{Sec-4}

As introduced in Section \ref{Sec-2}, a PQ symplectic transformation is given
by the action of the corresponding operator $\hat{\mathcal{O}}_{PQ}$ (\ref%
{O-PQ}). On the charge state (\ref{Q-state}), the action (\ref{Q-PQ}) can be
explicated as follows:\vspace{12pt}\begin{footnotesize}
\begin{eqnarray}
|\mathcal{Q}_{PQ}\rangle &:&=\hat{\mathcal{O}}_{PQ}|\mathcal{Q}\rangle
=p^{0}|0\rangle +\left( q_{0}+\rho p^{0}+\sum_{i=1}^{n}c_{i}p^{i}\right)
|1\rangle +\sum_{i=1}^{n}p^{i}|2^{i}\rangle +\sum_{i=1}^{n}\left(
q_{i}+c_{i}p^{0}+\sum_{j=1}^{n}p^{j}\Theta _{ji}\right) |3^{i}\rangle \notag
\\
&=&p^{0}|0\rangle +\tilde{q}_{0}|1\rangle +\sum_{i=1}^{n}p^{i}|2^{i}\rangle
+\sum_{i=1}^{n}\tilde{q}_{i}|3^{i}\rangle ,
\end{eqnarray}%
\end{footnotesize}
where:
\begin{eqnarray}
\tilde{q}_{0} &:&=q_{0}+\rho p^{0}+\sum_{i=1}^{n}c_{i}p^{i}, \\
\tilde{q}_{i} &:&=q_{i}+c_{i}p^{0}+\sum_{j=1}^{n}p^{j}\Theta _{ji}.
\end{eqnarray}%

As a consequence, PQ operators leave the EBH magnetic charges invariant,
while changing the electric ones~\cite{PQ}:%
\begin{equation}
\left\{
\begin{array}{l}
p^{0}\rightarrow p^{0}, \\
{p}^{i}\rightarrow p^{i} \\
q_{0}\rightarrow q_{0}+\rho p^{0}+\sum_{i=1}^{n}c_{i}p^{i}, \\
{q}_{i}\rightarrow q_{i}+c_{i}p^{0}+\sum_{j=1}^{n}p^{j}\Theta _{ji}.%
\end{array}%
\right.
\end{equation}

Let us now analyze the effects of the application of $\hat{\mathcal{O}}_{PQ}$
on the various $U$-orbit representatives/FTS elements presented in\ Section \ref%
{Sec-3} (considering before the ``large'' and then the
``small'' ones).

\begin{description}
\item[{{1.1]}}] \textbf{\textquotedblleft Large" KK configuration} :%
\begin{eqnarray}
\hat{\mathcal{O}}_{PQ}|\mathcal{Q}_{KK}\rangle &=&|\mathcal{Q}_{E}^{\prime
}\rangle ; \notag \\
|\mathcal{Q}_{E}^{\prime }\rangle &:&=p^{0}|0\rangle +\left( q_{0}+\rho
p^{0}\right) |1\rangle +p^{0}c_{i}|3^{i}\rangle. \label{11}
\end{eqnarray}%
Consequently, after a PQ transformation, the original KK dyonic EBH state is
changed to an~electric state. The value of the quartic invariant, and thus
of the Bekenstein--Hawking EBH entropy, generally changes:%
\begin{equation}
\mathcal{I}_{4}\left( \mathcal{Q}_{KK}\right) =-\left( p^{0}\right)
^{2}q_{0}^{2}\longrightarrow \mathcal{I}_{4}\left( \tilde{\mathcal{Q}}%
_{KK}\right) =-\left( p^{0}\right) ^{2}\left( q_{0}+\rho p^{0}\right) ^{2}-%
\frac{2}{3}\left( p^{0}\right) ^{4}\sum_{i,j,k=1}^{n}d^{ijk}c_{i}c_{j}c_{k}
\end{equation}%
Thus, depending on whether:%
\begin{equation}
\frac{2}{3}\sum_{i,j,k=1}^{n}d^{ijk}c_{i}c_{j}c_{k}\gtreqless -\left( \frac{%
q_{0}}{p^{0}}+\rho \right) ^{2}, \label{condd-1}
\end{equation}
\end{description}

\begin{enumerate}
\item
A \textquotedblleft large\textquotedblright\ ($\mathcal{I}_{4}>0$ : BPS or
non-BPS $Z_{H}=0$), a \textquotedblleft small\textquotedblright\ ($\mathcal{I%
}_{4}=0$ : BPS or non-BPS), or a \textquotedblleft large\textquotedblright\
non-BPS $Z_{H}\neq 0$ ($\mathcal{I}_{4}<0$) BH charge configuration is
generated by the action of the Peccei-Quinn symplectic group. Equation (\ref{condd-1}%
) shows that the relations among the components of $\mathcal{Q}$ and the
parameters of the Peccei--Quinn symplectic transformation turn out to be
crucial for the properties of the resulting charge configuration.
\end{enumerate}

\begin{enumerate}
\item [{{2.1]}}] \textbf{\textquotedblleft Large" electric configuration :}
\begin{eqnarray}
\hat{\mathcal{O}}_{PQ}|\mathcal{Q}_{E}\rangle &=&p^{0}|0\rangle +\rho
p^{0}|1\rangle +\sum_{i}\left( q_{i}+p^{0}c_{i}\right) |3^{i}\rangle =|%
\mathcal{Q}_{E}^{\prime \prime }\rangle +|\mathcal{Q}_{\rho eKK}\rangle , \\
|\mathcal{Q}_{E}^{\prime \prime }\rangle &:&=p^{0}|0\rangle +\sum_{i}\left(
q_{i}+p^{0}c_{i}\right) |3^{i}\rangle ; \\
|\mathcal{Q}_{\rho eKK}\rangle &:&=\rho p^{0}|1\rangle.
\end{eqnarray}%
Correspondingly, the quartic invariant changes as follows:%
\begin{equation}
\mathcal{I}_{4}\left( \mathcal{Q}_{E}\right) =-4p^{0}\mathcal{I}%
_{3}(q)\longrightarrow \mathcal{I}_{4}\left( \mathcal{Q}_{E}^{\prime \prime
}+\mathcal{Q}_{\rho eKK}\right) =-\rho ^{2}\left( p^{0}\right) ^{4}-4p^{0}%
\mathcal{I}_{3}(q_{i}+p^{0}c_{i}),
\end{equation}%
and there are three possibilities:%
\begin{eqnarray}
\mathbf{i} &:&\mathcal{I}_{4}\left( \mathcal{Q}_{E}^{\prime \prime }+%
\mathcal{Q}_{\rho eKK}\right) >0; \\
\mathbf{ii} &:&\mathcal{I}_{4}\left( \mathcal{Q}_{E}^{\prime \prime }+%
\mathcal{Q}_{\rho eKK}\right) <0; \\
\mathbf{iii} &\mathbf{:}&\mathcal{I}_{4}\left( \mathcal{Q}_{E}^{\prime
\prime }+\mathcal{Q}_{\rho eKK}\right) =0.
\end{eqnarray}%
In Cases $\mathbf{i}$
 and $\mathbf{ii}$, the resulting FTS element has rank four
, and thus, it corresponds to a ``large'' EBH, whereas in
Case $\mathbf{iii}$, the resulting FTS has rank three, corresponding to a
``small'' EBH. Therefore, depending on the sign of $p^{0}%
\mathcal{I}_{3}(q)$, the action of $\hat{\mathcal{O}}_{PQ}$ may result in
various types of {orbit transmutations}: a change of orbit
representative within the same $U$-orbit (at most, generally with a change
of value of $\mathcal{I}_{4}$) {or} a change of $U$-orbit, with
preservation of the FTS rank $=4$ (namely, a change of sign, and generally, of value of $\mathcal{I}_{4}$), {or}~(in Case $\mathbf{iii}$) a
change from a ``large'' $U$-orbit (rank-four) to a
``small'' $U$-orbit (rank-three).

\item[{{3.1]}}] \textbf{\textquotedblleft Large" magnetic configuration :}
\begin{equation}
\hat{\mathcal{O}}_{PQ}|\mathcal{Q}_{M}\rangle =\left(
q_{0}+\sum_{i}^{n}c_{i}p^{i}\right) |1\rangle
+\sum_{i}^{n}p^{i}|2^{i}\rangle +\sum_{i,j}^{n}p^{j}\Theta
_{ji}|3^{i}\rangle =|\mathcal{\tilde{Q}}_{M}\rangle +|\mathcal{\check{Q}}%
_{E}\rangle ,
\end{equation}%
where:%
\begin{eqnarray}
|\mathcal{\tilde{Q}}_{M}\rangle &:&=\tilde{q}_{0}|1\rangle
+\sum_{i}^{n}p^{i}|2^{i}\rangle , \label{Q-tilde-M} \\
\tilde{q}_{0} &:&=q_{0}+\sum_{i}^{n}c_{i}p^{i},
\end{eqnarray}%
and:%
\begin{eqnarray}
|\mathcal{\check{Q}}_{E}\rangle &:&=\sum_{i}^{n}\tilde{q}_{i}|3^{i}\rangle ,
\label{Q-hat-E} \\
\tilde{q}_{i} &:&=\sum_{j}^{n}p^{j}\Theta _{ji}. \label{Q-hat-E-2}
\end{eqnarray}%
In particular, $|\mathcal{\check{Q}}_{E}\rangle $ can correspond to one of
the three subcases {({2.1]}, {2.2]} and {2.3]}}) of Point
{{2}} in the Section~\ref{Sec-3}: 
\begin{gather}
\text{rank}\left( \mathcal{\check{Q}}_{E}\right) =3\Leftrightarrow
\sum_{i,j,k=1}^{n}d^{ijk}\tilde{q}_{i}\tilde{q}_{j}\tilde{q}_{k}\neq 0 \\
\Downarrow \notag \\
\mathcal{I}_{4}\left( \mathcal{\tilde{Q}}_{M}+\mathcal{\check{Q}}_{E}\right)
=-\left( \sum_{i=1}^{n}p^{i}\tilde{q}_{i}\right) ^{2}+4\tilde{q}_{0}\mathcal{%
I}_{3}(p)+4\{\mathcal{I}_{3}(p),\mathcal{I}_{3}(\tilde{q})\},
\end{gather}%
and there are three possibilities:%
\begin{eqnarray}
\mathbf{i} &:&\mathcal{I}_{4}\left( \mathcal{\tilde{Q}}_{M}+\mathcal{\check{Q%
}}_{E}\right) >0; \\
\mathbf{ii} &:&\mathcal{I}_{4}\left( \mathcal{\tilde{Q}}_{M}+\mathcal{\check{%
Q}}_{E}\right) <0; \\
\mathbf{iii} &\mathbf{:}&\mathcal{I}_{4}\left( \mathcal{\tilde{Q}}_{M}+%
\mathcal{\check{Q}}_{E}\right) =0.
\end{eqnarray}%
In Cases $\mathbf{i}$ and $\mathbf{ii}$, the resulting FTS element has rank four, and thus, it corresponds to a ``large'' EBH, whereas in
Case $\mathbf{iii}$, the resulting FTS has rank three, corresponding to a
``small'' EBH. Thus, depending on the sign of $q_{0}\mathcal{I%
}_{3}(p)$, the action of $\hat{\mathcal{O}}_{PQ}$ may once again result in
various types of {orbit transmutations}: a change of orbit
representative within the same $U$-orbit (at most, generally with a change
of value of $\mathcal{I}_{4}$), {or} a change of $U$-orbit, with
preservation of the FTS rank $=4$ (namely, a change of sign, and generally, of value of $\mathcal{I}_{4}$), or~(in Case $\mathbf{iii}$) a change from a
``large'' $U$-orbit (rank-four) to a ``small'' $%
U$-orbit (rank-three).%
\begin{gather}
\text{rank}\left( \mathcal{\check{Q}}_{E}\right) =2\Leftrightarrow
\sum_{i,j,k=1}^{n}d^{ijk}\tilde{q}_{i}\tilde{q}_{j}\tilde{q}_{k}=0,\text{%
~but~}\sum_{j,k=1}^{n}d^{ijk}\tilde{q}_{j}\tilde{q}_{k}\neq 0\text{~for~%
{at~least~}some~}i, \\
\Downarrow \notag \\
\mathcal{I}_{4}\left( \mathcal{\tilde{Q}}_{M}+\mathcal{\check{Q}}_{E}\right)
=-\left( \sum_{i=1}^{n}p^{i}\tilde{q}_{i}\right) ^{2}+4\tilde{q}_{0}\mathcal{%
I}_{3}(p)+4\{\mathcal{I}_{3}(p),\mathcal{I}_{3}(\tilde{q})\},
\end{gather}%
and there are three possibilities ($\mathbf{i}$--$\mathbf{iii}$) as above
(note that the term $\{\mathcal{I}_{3}(p),\mathcal{I}_{3}(\tilde{q})\}$ may
or may not be~vanishing).%
\begin{gather}
\text{rank}\left( \mathcal{\check{Q}}_{E}\right) =1\Leftrightarrow
\sum_{j,k=1}^{n}d^{ijk}\tilde{q}_{j}\tilde{q}_{k}=0~\forall i,\text{~but~}%
\tilde{q}_{i}\neq 0\text{~for~{at~least~}some~}i, \\
\Downarrow \notag \\
\mathcal{I}_{4}\left( \mathcal{\tilde{Q}}_{M}+\mathcal{\check{Q}}_{E}\right)
=-\left( \sum_{i=1}^{n}p^{i}\tilde{q}_{i}\right) ^{2}+4\tilde{q}_{0}\mathcal{%
I}_{3}(p),
\end{gather}%
and there are three possibilities ($\mathbf{i}$--$\mathbf{iii}$) as above.

\item[{{2.2]}}] \textbf{\textquotedblleft Small" rank-}$3$\textbf{\ electric
configuration} ($\mathcal{I}_{3}(q)\neq 0$) :
\begin{equation}
\hat{\mathcal{O}}_{PQ}|\mathcal{Q}_{3E}\rangle
=\sum_{i=1}^{n}q_{i}|3^{i}\rangle =|\mathcal{Q}_{3E}\rangle.
\end{equation}

\item[{{3.2]}}] \textbf{\textquotedblleft Small" rank-}$3$\textbf{\ magnectic
configuration }($\mathcal{I}_{3}(p)\neq 0$) :
\begin{equation}
\hat{\mathcal{O}}_{PQ}|\mathcal{Q}_{3M}\rangle
=\sum_{i=1}^{n}c_{i}p^{i}|1\rangle +\sum_{i=1}^{n}p^{i}|2^{i}\rangle
+\sum_{i,j=1}^{n}p^{j}\Theta _{ji}|3^{i}\rangle =\left. |\mathcal{\tilde{Q}}%
_{M}\rangle \right\vert _{q_{0}=0}+|\mathcal{\check{Q}}_{E}\rangle ,
\end{equation}%
where $\left. |\mathcal{\tilde{Q}}_{M}\rangle \right\vert _{q_{0}=0}$ is the
$q_{0}=0$ case limit of $|\mathcal{\tilde{Q}}_{M}\rangle $ defined in Equation (\ref%
{Q-tilde-M}), and $|\mathcal{\check{Q}}_{E}\rangle $ is given in Equations (\ref%
{Q-hat-E}) and (\ref{Q-hat-E-2}). There~are three subcases, respectively
corresponding to rank $\left( \mathcal{\check{Q}}_{E}\right) =3,2,1$.%
\begin{gather}
\text{rank}\left( \mathcal{\check{Q}}_{E}\right) =3\Leftrightarrow
\sum_{i,j,k=1}^{n}d^{ijk}\tilde{q}_{i}\tilde{q}_{j}\tilde{q}_{k}\neq 0 \\
\Downarrow \notag \\
\mathcal{I}_{4}\left( \left. \mathcal{\tilde{Q}}_{M}\right\vert _{q_{0}=0}+%
\mathcal{\check{Q}}_{E}\right) =-\left( \sum_{i=1}^{n}p^{i}\tilde{q}%
_{i}\right) ^{2}+4\sum_{i=1}^{n}c_{i}p^{i}\mathcal{I}_{3}(p)+4\{\mathcal{I}%
_{3}(p),\mathcal{I}_{3}(\tilde{q})\},
\end{gather}%
and there are three possibilities:%
\begin{eqnarray}
\mathbf{i} &:&\mathcal{I}_{4}\left( \left. \mathcal{\tilde{Q}}%
_{M}\right\vert _{q_{0}=0}+\mathcal{\check{Q}}_{E}\right) >0; \\
\mathbf{ii} &:&\mathcal{I}_{4}\left( \left. \mathcal{\tilde{Q}}%
_{M}\right\vert _{q_{0}=0}+\mathcal{\check{Q}}_{E}\right) <0; \\
\mathbf{iii} &\mathbf{:}&\mathcal{I}_{4}\left( \left. \mathcal{\tilde{Q}}%
_{M}\right\vert _{q_{0}=0}+\mathcal{\check{Q}}_{E}\right) =0.
\end{eqnarray}%
In Cases $\mathbf{i}$ and $\mathbf{ii}$, the resulting FTS element has rank four, and thus, it corresponds to a ``large'' EBH, whereas in
Case $\mathbf{iii}$, the resulting FTS has rank three, corresponding to a
``small'' EBH. Thus, in~this case, the action of $\hat{%
\mathcal{O}}_{PQ}$ may once again result in various types of {orbit
transmutations}: a change from a~``small'' $U$-orbit (rank-three) to a ``large'' $U$-orbit (rank-four, with positive or
negative $\mathcal{I}_{4}$), {or}~(in~Case~$\mathbf{iii}$) a
preservation of the rank$=3$ of the original FTS element, but generally with
a change of orbit representative (depending on the real form of the theory
under consideration, this may result necessarily in remaining in the same
rank-three orbit, as in $\mathcal{N}=8$ supergravity, or possibly in
switching to another rank-three orbit, as in $\mathcal{N}=2$ supergravity
with symmetric, very special K\"{a}hler vector multiplets' scalar manifolds \cite{Marrani-review, Small-Orbits}).%
\begin{gather}
\text{rank}\left( \mathcal{\check{Q}}_{E}\right) =2\Leftrightarrow
\sum_{i,j,k=1}^{n}d^{ijk}\tilde{q}_{i}\tilde{q}_{j}\tilde{q}_{k}=0,\text{%
~but~}\sum_{j,k=1}^{n}d^{ijk}\tilde{q}_{j}\tilde{q}_{k}\neq 0\text{~for~%
{at~least~}some~}i, \\
\Downarrow \notag \\
\mathcal{I}_{4}\left( \left. \mathcal{\tilde{Q}}_{M}\right\vert _{q_{0}=0}+%
\mathcal{\check{Q}}_{E}\right) =-\left( \sum_{i=1}^{n}p^{i}\tilde{q}%
_{i}\right) ^{2}+4\sum_{i=1}^{n}c_{i}p^{i}\mathcal{I}_{3}(p)+4\{\mathcal{I}%
_{3}(p),\mathcal{I}_{3}(\tilde{q})\},
\end{gather}%
and there are three possibilities $\mathbf{i}$--$\mathbf{iii}$ as above (note
that the term $\{\mathcal{I}_{3}(p),\mathcal{I}_{3}(\tilde{q})\}$ may or may
not be~vanishing).%
\begin{gather}
\text{rank}\left( \mathcal{\check{Q}}_{E}\right) =1\Leftrightarrow
\sum_{j,k=1}^{n}d^{ijk}\tilde{q}_{j}\tilde{q}_{k}=0~\forall i,\text{~but~}%
\tilde{q}_{i}\neq 0\text{~for~{at~least~}some~}i, \\
\Downarrow \notag \\
\mathcal{I}_{4}\left( \left. \mathcal{\tilde{Q}}_{M}\right\vert _{q_{0}=0}+%
\mathcal{\check{Q}}_{E}\right) =-\left( \sum_{i=1}^{n}p^{i}\tilde{q}%
_{i}\right) ^{2}+4\sum_{i=1}^{n}c_{i}p^{i}\mathcal{I}_{3}(p),
\end{gather}%
and there are three possibilities $\mathbf{i}$--$\mathbf{iii}$ as above.

\item[{{2.3]}}] \textbf{"Small" rank-}$\mathbf{2}$\textbf{\ electric
configuration} ($\mathcal{I}_{3}(q)=0$, but $\partial \mathcal{I}%
_{3}(q)/\partial q_{i}\neq 0$ \textit{at least} for some $i$) :
\begin{equation}
\hat{\mathcal{O}}_{PQ}|\mathcal{Q}_{2E}\rangle
=\sum_{i=1}^{n}q_{i}|3^{i}\rangle =|\mathcal{Q}_{2E}\rangle.
\end{equation}

\item[{{3.3]}}] \textbf{\textquotedblleft Small" rank-}$\mathbf{2}$\textbf{\
magnetic configuration} ($\mathcal{I}_{3}(p)=0$, but $\partial \mathcal{I}%
_{3}(p)/\partial p^{i}\neq 0$ \textit{at least} for some $i$) :
\begin{eqnarray}
\hat{\mathcal{O}}_{PQ}|\mathcal{Q}_{2M}\rangle
&=&\sum_{i=1}^{n}c_{i}p^{i}|1\rangle +\sum_{i=1}^{n}p^{i}|2^{i}\rangle
+\sum_{i,j=1}^{n}p^{j}\Theta _{ji}|3^{i}\rangle \\
&=&|\mathcal{\mathring{Q}}_{2M}\rangle +|\mathcal{\check{Q}}_{E}\rangle ,
\end{eqnarray}%
where:%
\begin{equation}
|\mathcal{\mathring{Q}}_{2M}\rangle:=\sum_{i=1}^{n}c_{i}p^{i}|1\rangle
+\sum_{i=1}^{n}p^{i}|2^{i}\rangle.
\end{equation}%
Depending on rank $\left( \mathcal{\check{Q}}_{E}\right) =3,2,1$, one has
still three subcases.%
\begin{gather}
\text{rank}\left( \mathcal{\check{Q}}_{E}\right) =3\Leftrightarrow
\sum_{i,j,k=1}^{n}d^{ijk}\tilde{q}_{i}\tilde{q}_{j}\tilde{q}_{k}\neq 0 \\
\Downarrow \notag \\
\mathcal{I}_{4}\left( \mathcal{\mathring{Q}}_{2M}+\mathcal{\check{Q}}%
_{E}\right) =-\left( \sum_{i=1}^{n}p^{i}\tilde{q}_{i}\right) ^{2}+4\{%
\mathcal{I}_{3}(p),\mathcal{I}_{3}(\tilde{q})\},
\end{gather}%
and there are three possibilities:%
\vspace{12pt}
\begin{eqnarray}
\mathbf{i} &:&\mathcal{I}_{4}\left( \mathcal{\mathring{Q}}_{2M}+\mathcal{%
\check{Q}}_{E}\right) >0; \\
\mathbf{ii} &:&\mathcal{I}_{4}\left( \mathcal{\mathring{Q}}_{2M}+\mathcal{%
\check{Q}}_{E}\right) <0; \\
\mathbf{iii} &\mathbf{:}&\mathcal{I}_{4}\left( \mathcal{\mathring{Q}}_{2M}+%
\mathcal{\check{Q}}_{E}\right) =0.
\end{eqnarray}%
In Cases $\mathbf{i}$ and $\mathbf{ii}$, the resulting FTS element has rank four, and thus, it corresponds to a ``large'' EBH, whereas in
Case $\mathbf{iii}$, the resulting FTS has rank three, corresponding to a
``small'' EBH. Thus, in this case, the action of $\hat{%
\mathcal{O}}_{PQ}$ may once again result in various types of {orbit
transmutations}: a change from a~``small'' $U$-orbit (rank-two) to a~``large'' $U$-orbit (rank-four, with positive or
negative $\mathcal{I}_{4}$), {{or}} (in Case $\mathbf{iii}$) a~change
from a ``small'' rank-two orbit to a~``small''
rank-three orbit.%
\begin{gather}
\text{rank}\left( \mathcal{\check{Q}}_{E}\right) =2\Leftrightarrow
\sum_{i,j,k=1}^{n}d^{ijk}\tilde{q}_{i}\tilde{q}_{j}\tilde{q}_{k}=0,\text{%
~but~}\sum_{j,k=1}^{n}d^{ijk}\tilde{q}_{j}\tilde{q}_{k}\neq 0\text{~for~%
{at~least~}some~}i, \\
\Downarrow \notag \\
\mathcal{I}_{4}\left( \mathcal{\mathring{Q}}_{2M}+\mathcal{\check{Q}}%
_{E}\right) =-\left( \sum_{i=1}^{n}p^{i}\tilde{q}_{i}\right) ^{2}+4\{%
\mathcal{I}_{3}(p),\mathcal{I}_{3}(\tilde{q})\},
\end{gather}%
and there are three possibilities (note that the term $\{\mathcal{I}_{3}(p),%
\mathcal{I}_{3}(\tilde{q})\}$ may or may not be vanishing):%
\begin{eqnarray}
\mathbf{i} &:&\mathcal{I}_{4}\left( \mathcal{\mathring{Q}}_{2M}+\mathcal{%
\check{Q}}_{E}\right) >0; \\
\mathbf{ii} &:&\mathcal{I}_{4}\left( \mathcal{\mathring{Q}}_{2M}+\mathcal{%
\check{Q}}_{E}\right) <0; \\
\mathbf{iii} &\mathbf{:}&\mathcal{I}_{4}\left( \mathcal{\mathring{Q}}_{2M}+%
\mathcal{\check{Q}}_{E}\right) =0.
\end{eqnarray}%
Once again, in Cases $\mathbf{i}$ and $\mathbf{ii}$, the resulting FTS
element has rank four, and thus, it corresponds to a~``large''
EBH, whereas in Case $\mathbf{iii}$, the resulting FTS has one of the
possible ranks $3, 2, 1$, corresponding to a ``small'' EBH.
Thus, in this case that the action of $\hat{\mathcal{O}}_{PQ}$ may once again
result in various types of {orbit transmutations}: a change from a
``small'' $U$-orbit (rank-two) to a ``large'' $%
U$-orbit (rank-four, with~positive or negative $\mathcal{I}_{4}$), {{or}}
(in Case $\mathbf{iii}$) the change from a ``small'' $U$%
-orbit (rank-two) to ``small'' orbit, of~possible rank 3, $2$ %
,$1$. In the case of rank$=2$, generally a change of orbit representative 
takes place (as above, depending on the real form of the theory under
consideration, this may result necessarily in remaining in the same rank-two orbit, or possibly in switching to another rank-two orbit~\cite%
{Marrani-review, Small-Orbits}).%
\begin{gather}
\text{rank}\left( \mathcal{\check{Q}}_{E}\right) =1\Leftrightarrow
\sum_{j,k=1}^{n}d^{ijk}\tilde{q}_{j}\tilde{q}_{k}=0~\forall i,\text{~but~}%
\tilde{q}_{i}\neq 0\text{~for~{at~least~}some~}i, \\
\Downarrow \notag \\
\mathcal{I}_{4}\left( \mathcal{\mathring{Q}}_{2M}+\mathcal{\check{Q}}%
_{E}\right) =-\left( \sum_{i=1}^{n}p^{i}\tilde{q}_{i}\right) ^{2},
\end{gather}%
and there are two possibilities:%
\begin{eqnarray}
\mathbf{i} &:&\mathcal{I}_{4}\left( \mathcal{\mathring{Q}}_{2M}+\mathcal{%
\check{Q}}_{E}\right) <0; \\
\mathbf{ii} &:&\mathcal{I}_{4}\left( \mathcal{\mathring{Q}}_{2M}+\mathcal{%
\check{Q}}_{E}\right) =0.
\end{eqnarray}%
In Case $\mathbf{i}$, the resulting FTS element has rank four (with negative $%
\mathcal{I}_{4}$), and thus, it corresponds to a~``large''
EBH, whereas in Case $\mathbf{ii}$, the resulting FTS has one of the possible
ranks $3,2,1$, corresponding to a~``small'' EBH. Thus, in
this case, the action of $\hat{\mathcal{O}}_{PQ}$ may once again result in
various types of {orbit transmutations}: a change from a
``small'' $U$-orbit (rank-two) to a ``large'' $%
U$-orbit (rank-four, with negative $\mathcal{I}_{4}$), {or} (in Case $%
\mathbf{ii}$) the change from a ``small'' $U$-orbit (rank-two%
) to ``small'' orbit, of possible rank 3, $2$, $1$. For the 
case of rank$=2$, generally a change of orbit representative takes place (as
before, depending on the real form of the theory under consideration, this
may result necessarily in remaining at the same rank-two orbit, or
possibly in switching to another rank-two orbit \cite{Marrani-review,
Small-Orbits}).

\item[{{2.4]}}] \textbf{\textquotedblleft Small" rank-}$1$\textbf{\ electric
configuration} ($\partial \mathcal{I}_{3}(q)/\partial q_{i}=0~\forall i$,
but $q_{i}\neq 0$ \textit{at least} for some $i$) :
\begin{equation}
\hat{\mathcal{O}}_{PQ}|\mathcal{Q}_{1E}\rangle
=\sum_{i=1}^{n}q_{i}|3^{i}\rangle =|\mathcal{Q}_{1E}\rangle.
\end{equation}

\item[{{3.4]}}] \textbf{\textquotedblleft Small" rank-}$\mathbf{1}$\textbf{\
magnetic configuration} ($\partial \mathcal{I}_{3}(p)/\partial
p^{i}=0~\forall i$, but $p^{i}\neq 0$ \textit{at least} for some $i$) :
\begin{equation}
\hat{\mathcal{O}}_{PQ}|\mathcal{Q}_{1M}\rangle
=\sum_{i=1}^{n}c_{i}p^{i}|1\rangle +\sum_{i=1}^{n}p^{i}|2^{i}\rangle
+\sum_{i,j=1}^{n}p^{j}\Theta _{ji}|3^{i}\rangle =|\mathcal{\mathring{Q}}%
_{1M}\rangle +|\mathcal{\check{Q}}_{E}\rangle ,
\end{equation}%
where:%
\begin{equation}
|\mathcal{\mathring{Q}}_{1M}\rangle:=\sum_{i=1}^{n}c_{i}p^{i}|1\rangle
+\sum_{i=1}^{n}p^{i}|2^{i}\rangle.
\end{equation}%
{D}epending on rank$\left( \mathcal{\check{Q}}_{E}\right) =3,2,1$,
one has still three subcases.%
\begin{gather}
\text{rank}\left( \mathcal{\check{Q}}_{E}\right) =3\Leftrightarrow
\sum_{i,j,k=1}^{n}d^{ijk}\tilde{q}_{i}\tilde{q}_{j}\tilde{q}_{k}\neq 0 \\
\Downarrow \notag \\
\mathcal{I}_{4}\left( \mathcal{\mathring{Q}}_{1M}+\mathcal{\check{Q}}%
_{E}\right) =-\left( \sum_{i=1}^{n}p^{i}\tilde{q}_{i}\right) ^{2},
\end{gather}%
and there are two possibilities:%
\begin{eqnarray}
\mathbf{i} &:&\mathcal{I}_{4}\left( \mathcal{\mathring{Q}}_{1M}+\mathcal{%
\check{Q}}_{E}\right) <0; \\
\mathbf{ii} &:&\mathcal{I}_{4}\left( \mathcal{\mathring{Q}}_{1M}+\mathcal{%
\check{Q}}_{E}\right) =0.
\end{eqnarray}%
In Case $\mathbf{i}$ the resulting FTS element has rank four (with negative $%
\mathcal{I}_{4}$), and thus, it corresponds to a ``large''
EBH, whereas in Case $\mathbf{ii}$, the resulting FTS has one of the possible
ranks $3,2,1$, corresponding to a~``small'' EBH. Thus, in
this case, the action of $\hat{\mathcal{O}}_{PQ}$ results in various types of
{orbit transmutations}: a~change from a ``small'' $U$%
-orbit (rank-one) to a ``large'' $U$-orbit (rank-four, with
positive or negative $\mathcal{I}_{4}$), {or} (in Case $\mathbf{ii}$)
the change from a~``small'' $U$-orbit (rank-one) to
``small'' orbit, of possible rank 3, $2$, $1$; in~the case of
rank$=1$, generally a~change of orbit representative takes place, and
regardless of the real form of the theory under consideration, one~necessarily remains in the same rank-one orbit, which is always unique
\cite{Marrani-review, Small-Orbits}.%
\begin{gather}
\text{rank}\left( \mathcal{\check{Q}}_{E}\right) =2\Leftrightarrow
\sum_{i,j,k=1}^{n}d^{ijk}\tilde{q}_{i}\tilde{q}_{j}\tilde{q}_{k}=0,\text{%
~but~}\sum_{j,k=1}^{n}d^{ijk}\tilde{q}_{j}\tilde{q}_{k}\neq 0\text{~for~%
{at~least~}some~}i, \\
\Downarrow \notag \\
\mathcal{I}_{4}\left( \mathcal{\mathring{Q}}_{1M}+\mathcal{\check{Q}}%
_{E}\right) =-\left( \sum_{i=1}^{n}p^{i}\tilde{q}_{i}\right) ^{2},
\end{gather}%
and, as above, there are two possibilities:%
\begin{eqnarray}
\mathbf{i} &:&\mathcal{I}_{4}\left( \mathcal{\mathring{Q}}_{1M}+\mathcal{%
\check{Q}}_{E}\right) <0; \\
\mathbf{ii} &:&\mathcal{I}_{4}\left( \mathcal{\mathring{Q}}_{1M}+\mathcal{%
\check{Q}}_{E}\right) =0.
\end{eqnarray}%
In Case $\mathbf{i}$, the resulting FTS element has rank four, and thus, it
corresponds to a ``large'' EBH, whereas~in Case $\mathbf{ii}$,
the resulting FTS has possible ranks $3,2,1$, corresponding to a
``small'' EBH. Thus, in~this case, the action of $\hat{%
\mathcal{O}}_{PQ}$ results in various types of {orbit transmutations}: a change from a ``small'' $U$-orbit (rank-one) to a
``large'' $U$-orbit (rank-four, with negative $\mathcal{I}_{4}$%
), {or} (in Case $\mathbf{ii}$) the change from a~``small'' $U$-orbit (rank-one) to ``small'' orbit, of possible
rank 3, $2$, $1$; in the case of rank$=1$, once~again, one~necessarily
remains in the same rank-one orbit~\cite{Marrani-review, Small-Orbits}.%
\begin{gather}
\text{rank}\left( \mathcal{\check{Q}}_{E}\right) =1\Leftrightarrow
\sum_{j,k=1}^{n}d^{ijk}\tilde{q}_{j}\tilde{q}_{k}=0~\forall i,\text{~but~}%
\tilde{q}_{i}\neq 0\text{~for~{at~least~}some~}i, \\
\Downarrow \notag \\
\mathcal{I}_{4}\left( \mathcal{\mathring{Q}}_{1M}+\mathcal{\check{Q}}%
_{E}\right) =-\left( \sum_{i=1}^{n}p^{i}\tilde{q}_{i}\right) ^{2},
\end{gather}%
and there are two possibilities:%
\begin{eqnarray}
\mathbf{i} &:&\mathcal{I}_{4}\left( \mathcal{\mathring{Q}}_{1M}+\mathcal{%
\check{Q}}_{E}\right) <0; \\
\mathbf{ii} &:&\mathcal{I}_{4}\left( \mathcal{\mathring{Q}}_{1M}+\mathcal{%
\check{Q}}_{E}\right) =0.
\end{eqnarray}%
In Case $\mathbf{i}$, the resulting FTS element has rank four (with negative $%
\mathcal{I}_{4}$), and thus, it corresponds to a~``large''
EBH, whereas in Case $\mathbf{ii}$, the resulting FTS has possible rank $3,2,1
$, corresponding to a~``small'' EBH. Thus, in~this case, the
action of $\hat{\mathcal{O}}_{PQ}$ results in various types of {orbit
transmutations}: a~change from a~``small'' $U$-orbit (rank-one) to a ``large'' $U$-orbit (rank-four, with negative $%
\mathcal{I}_{4}$), {or}~(in~Case~$\mathbf{ii}$), the change from a
``small'' $U$-orbit (rank-one) to ``small''
orbit, of possible rank 3, $2$, $1$; in the case of rank$=1$, once~again,
one~necessarily remains in the same rank-one orbit \cite{Marrani-review,
Small-Orbits}.

\item[{{1.3]}}] 
\textbf{\textquotedblleft Small" rank-}$\mathbf{1}$\textbf{\
magnetic KK configuration} :
\begin{equation}
\hat{\mathcal{O}}_{PQ}|\mathcal{Q}_{mKK}\rangle =p^{0}|0\rangle +\rho
p^{0}|1\rangle +p^{0}\sum_{i=1}^{n}c_{i}|3^{i}\rangle =\left. |\mathcal{Q}%
_{E}^{\prime }\rangle \right\vert _{q_{0}=0},
\end{equation}%
where $\left. |\mathcal{Q}_{E}^{\prime }\rangle \right\vert _{q_{0}=0}$ is
the $q_{0}=0$ limit of $|\mathcal{Q}_{E}^{\prime }\rangle $ given by (\ref%
{11}). Thus, in this case, the Peccei--Quinn symplectic transformation
generate{s} a $\rho $-dependent graviphoton electric charge and $c_{i}$%
-dependent electric charges.~These latter in Type II compactifications
correspond to a stack of $D2$ branes depending on the components of the
second Chern class $c_{2}$ of the Calabi--Yau three-fold. The~corresponding
transformation of I4
 reads:%
\begin{equation}
\mathcal{I}_{4}(\mathcal{Q}_{mKK})=0\longrightarrow \mathcal{I}_{4}(\left. |%
\mathcal{Q}_{E}^{\prime }\rangle \right\vert _{q_{0}=0})-\left( p^{0}\right)
^{4}\left( \rho ^{2}+\frac{2}{3}\sum_{i,j,k=1}^{n}d^{ijk}c_{i}c_{j}c_{k}%
\right) \gtreqless 0 \label{Jan}
\end{equation}%
Thus, depending on the sign of the r.h.s. of $\mathcal{I}_{4}(\left. |%
\mathcal{Q}_{E}^{\prime }\rangle \right\vert _{q_{0}=0})$, a
\textquotedblleft large\textquotedblright\ ($\mathcal{I}_{4}>0$: BPS, or
non-BPS $Z_{H}=0$), a \textquotedblleft small\textquotedblright\ ($\mathcal{I%
}_{4}=0$:BPS or non-BPS), or a \textquotedblleft large\textquotedblright\
non-BPS $Z_{H}\neq 0$ ($\mathcal{I}_{4}<0$) BH charge configuration is~generated.

\item[{{1.2]}}] \textbf{\textquotedblleft Small" rank-}$1$\textbf{\ electric KK
configuration} :
\begin{equation}
\hat{\mathcal{O}}_{PQ}|\mathcal{Q}_{eKK}\rangle =q_{0}|1\rangle =|\mathcal{Q}%
_{eKK}\rangle.
\end{equation}
\end{enumerate}

\section{Superpositions}\label{Sec-5}

In the Hilbert space representation, the superposition of EBH charge
configurations allows one to deal with the whole charge scenario for the
state configurations of EBHs. A quite general class of superpositions can be
written in the following form:
\begin{eqnarray}
|\mathbf{Q}\rangle &:&=\alpha _{1}|\mathcal{Q}_{KK}\rangle +\alpha _{2}|%
\mathcal{Q}_{E}\rangle +\alpha _{3}|\mathcal{Q}_{M}\rangle +\alpha _{4}|%
\mathcal{Q}_{3E}\rangle +\alpha _{5}|\mathcal{Q}_{3M}\rangle \notag \\
&+&\alpha _{6}|\mathcal{Q}_{2E}\rangle +\alpha _{7}|\mathcal{Q}_{2M}\rangle
+\alpha _{8}|\mathcal{Q}_{1E}\rangle +\alpha _{9}|\mathcal{Q}_{1E}\rangle
\notag \\
&+&\alpha _{10}|\mathcal{Q}_{mKK}\rangle +\alpha _{11}|\mathcal{Q}%
_{eKK}\rangle , \label{Q-bold}
\end{eqnarray}%
where $\alpha _{1},...,\alpha _{11}\in \mathbb{C}$ (see the next section for
a discussion of the necessary emergence of complex parameters). We can
rewrite this state as:
\begin{eqnarray}
|\mathbf{Q}\rangle &=&(\alpha _{1}+\alpha _{2}+\alpha _{10})p^{0}|0\rangle
+(\alpha _{1}+\alpha _{3}+\alpha _{11})q_{0}|1\rangle \notag \\
&+&(\alpha _{2}+\alpha _{4}+\alpha _{6}+\alpha
_{8})\sum_{i=1}^{n}q_{i}|3^{i}\rangle +(\alpha _{3}+\alpha _{5}+\alpha
_{7}+\alpha _{9})\sum_{i=1}^{n}p^{i}|2^{i}\rangle ,
\end{eqnarray}%
and the the explicit expression of the quartic invariant for such a
superposed EBH charge configuration~reads:
\begin{eqnarray}
\mathcal{I}_{4}(\mathbf{Q}) &=&-(\alpha _{1}+\alpha _{2}+\alpha
_{10})^{2}(\alpha _{1}+\alpha _{3}+\alpha _{11})^{2}\left( p^{0}q_{0}\right)
^{2} \notag \\
&&-(\alpha _{2}+\alpha _{4}+\alpha _{6}+\alpha _{8})^{2}(\alpha _{3}+\alpha
_{5}+\alpha _{7}+\alpha _{9})^{2}\sum_{i=1}^{n}\left( p^{i}q_{i}\right) ^{2}
\notag \\
&-&2(\alpha _{1}+\alpha _{2}+\alpha _{10})(\alpha _{1}+\alpha _{3}+\alpha
_{11})(\alpha _{2}+\alpha _{4}+\alpha _{6}+\alpha _{8})(\alpha _{3}+\alpha
_{5}+\alpha _{7}+\alpha _{9})p^{0}q_{0}\sum_{i=1}^{n}p^{i}q_{i} \notag \\
&+&4(\alpha _{1}+\alpha _{3}+\alpha _{11})(\alpha _{3}+\alpha _{5}+\alpha
_{7}+\alpha _{9})^{3}q_{0}\mathcal{I}_{3}(p) \notag \\
&&-4(\alpha _{1}+\alpha _{2}+\alpha _{10})(\alpha _{2}+\alpha _{4}+\alpha
_{6}+\alpha _{8})^{3}p^{0}\mathcal{I}_{3}(q) \notag \\
&&+4(\alpha _{3}+\alpha _{5}+\alpha _{7}+\alpha _{9})^{3}(\alpha _{2}+\alpha
_{4}+\alpha _{6}+\alpha _{8})^{3}\{\mathcal{I}_{3}(p),\mathcal{I}_{3}(q)\},
\end{eqnarray}%
and thus, the Bekenstein--Hawking EBH entropy is given by:%
\begin{equation}
S\left( \mathbf{Q}\right) =\pi \sqrt{\left\vert \mathcal{I}_{4}(\mathbf{Q}%
)\right\vert }.
\end{equation}

As a simple example, let us consider the superposition of a
``large'' KK and a ``large'' electric charge~configuration:
\begin{equation}
|\mathbf{Q}_{KK,E}\rangle:=\alpha _{1}|\mathcal{Q}_{KK}\rangle +\alpha _{2}|%
\mathcal{Q}_{E}\rangle ,~\alpha _{1},\alpha _{2}\in \mathbb{C},
\end{equation}%
yielding to
\begin{eqnarray}
\mathcal{I}_{4}(\mathbf{Q}_{KK,E}) &=&-(\alpha _{1}+\alpha _{2})^{2}\alpha
_{1}^{2}\left( p^{0}q_{0}\right) ^{2}-4(\alpha _{1}+\alpha _{2})\alpha
_{2}{}^{3}p^{0}\mathcal{I}_{3}(q), \\
S\left( \mathbf{Q}\right) &=&\pi \sqrt{\left\vert (\alpha _{1}+\alpha
_{2})^{2}\alpha _{1}^{2}\left( p^{0}q_{0}\right) ^{2}+4(\alpha _{1}+\alpha
_{2})\alpha _{2}{}^{3}p^{0}\mathcal{I}_{3}(q)\right\vert }.
\end{eqnarray}

\section{\label{Sec-6} Entanglement PQ Operators and Complexification}

{Generalized} PQ operators can be defined as tensor products of PQ
operators (\ref{O-PQ}). The simplest ({i.e.},~two-fold) generalized PQ
operator is defined as:
\begin{equation}
\hat{\mathcal{O}}_{PQ}^{(2)}:=\hat{\mathcal{O}}_{PQ}\otimes \hat{\mathcal{%
O^{\prime }}}_{PQ},
\end{equation}%
where (recalling Equation (\ref{O-PQ})):\begin{footnotesize}
\begin{eqnarray}
\hat{\mathcal{O}}_{PQ} &=&|0\rangle \langle 0|+|1\rangle \langle 1|+\rho
|1\rangle \langle 0|+\sum_{i=1}^{n}\left( |2^{i}\rangle \langle
2^{i}|+|3^{i}\rangle \langle 3^{i}|\right) +\sum_{i=1}^{n}\left(
c_{i}|1\rangle \langle 2^{i}|+c_{i}|3^{i}\rangle \langle 0|\right)
+\sum_{i,j=1}^{n}\Theta _{ij}|3^{i}\rangle \langle 2^{i}|, \\
\hat{\mathcal{O^{\prime }}}_{PQ} &=&|0\rangle \langle 0|+|1\rangle \langle
1|+\rho ^{\prime }|1\rangle \langle 0|+\sum_{i=1}^{n}\left( |2^{i}\rangle
\langle 2^{i}|+|3^{i}\rangle \langle 3^{i}|\right) +\sum_{i=1}^{n}\left(
c_{i}^{\prime }|1\rangle \langle 2^{i}|+c_{i}^{\prime }|3^{i}\rangle \langle
0|\right) +\sum_{i,j=1}^{n}\Theta _{ij}^{\prime }|3^{i}\rangle \langle
2^{i}|.
\end{eqnarray}
\end{footnotesize}

For instance, the action of $\hat{\mathcal{O}}_{PQ}^{(2)}$ (with $c_{i}=0$)
on a composed configuration of two KK ``large'' EBH states
(which can also be interpreted as the state of a two-centered EBH in which
each center is characterized by a KK ``large'' charge
configuration) is given by (recall Equation (\ref{11})):
\begin{eqnarray}
\left. \hat{\mathcal{O}}_{PQ}^{(2)}\right\vert _{c_{i}=0}\left( |\mathcal{Q}%
_{KK}\rangle \otimes |\mathcal{Q}_{KK}^{\prime }\rangle \right) &:&=\left.
\hat{\mathcal{O}}_{PQ}\right\vert _{c_{i}=0}|\mathcal{Q}_{KK}\rangle \otimes
\left. \hat{\mathcal{O^{\prime }}}_{PQ}\right\vert _{c_{i}=0}|\mathcal{Q}%
_{KK}^{\prime }\rangle =|\tilde{\mathcal{Q}}_{KK}\rangle \otimes |\tilde{%
\mathcal{Q}}_{KK}^{\prime }\rangle \notag \\
&=&\left( p^{0}|0\rangle +\left( q_{0}+\rho p^{0}\right) |1\rangle \right)
\otimes \left( p^{0\prime }|0\rangle +\left( q_{0}^{\prime }+\rho ^{\prime
}p^{0\prime }\right) |1\rangle \right) \notag \\
&=&p^{0}p^{0\prime }|00\rangle +p^{0}\left( q_{0}^{\prime }+\rho ^{\prime
}p^{0\prime }\right) |01\rangle \notag \\
&&+\left( q_{0}+\rho p^{0}\right) p^{0\prime }|10\rangle +\left( q_{0}+\rho
p^{0}\right) \left( q_{0}^{\prime }+\rho ^{\prime }p^{0\prime }\right)
|11\rangle ,
\end{eqnarray}%
where we introduced:%
\begin{equation}
|\alpha \beta \rangle:=|\alpha \rangle \otimes |\beta \rangle ,~\alpha
,\beta =0,1
\end{equation}%
as the basis of the Hilbert space of two ``large'' KK EBHs
(or, equivalently, of a two-centered EBH in which each center is
characterized by a KK ``large'' charge
configuration).\smallskip

However, in order to generate {entanglement} in EBH systems through
PQ symplectic transformations and thus realize ``EBH
quantum circuits'' and ``EBH quantum gates'' in the context of
BHQC \cite{prudencio1}, one~can employ simpler composed PQ operators, namely:
\begin{equation}
\hat{\mathcal{E}}_{PQ}:=\alpha \hat{\mathbb{I}}\otimes \hat{\mathcal{O}}%
_{PQ}+\hat{\mathcal{O}}_{PQ}\otimes \beta \hat{\mathbb{I}}, \label{fip}
\end{equation}%
which we name the (two-fold) {entangled} PQ operator; $\hat{\mathbb{I}}$
denotes the identity operator and $\alpha ,\beta \in \mathbb{C}$.

As an example, let us again consider a composed configuration of two KK
``large'' EBH states. The~action of $\hat{\mathcal{E}}_{PQ}$ (%
\ref{fip}) (with $c_{i}=0$) is given by:
\begin{eqnarray}
\left. \hat{\mathcal{E}}_{PQ}\right\vert _{c_{i}=0}\left( |\mathcal{Q}%
_{KK}\rangle \otimes |\mathcal{Q}_{KK}^{\prime }\rangle \right) &=&\alpha
\left( p^{0}|0\rangle +q_{0}|1\rangle \right) \otimes \left( p^{0\prime
}|0\rangle +\left( q_{0}^{\prime }+\rho p^{0\prime }\right) |1\rangle \right)
\notag \\
&+&\beta \left( p^{0}|0\rangle +\left( q_{0}+\rho p^{0}\right) |1\rangle
\right) \otimes p^{0\prime }|0\rangle +q_{0}^{\prime }|1\rangle \notag \\
&=&\alpha \left( p^{0}p^{0\prime }|00\rangle +p^{0}\left( q_{0}^{\prime
}+\rho p^{0\prime }\right) |01\rangle +q_{0}p^{0\prime }|10\rangle
+q_{0}\left( q_{0}^{\prime }+\rho p^{0\prime }\right) |11\rangle \right)
\notag \\
&&+\beta \left( p^{0}p^{0\prime }|00\rangle +p^{0}q_{0}^{\prime }|01\rangle
+\left( q_{0}+\rho p^{0}\right) p^{0\prime }|10\rangle +\left( q_{0}+\rho
p^{0}\right) q_{0}^{\prime }|11\rangle \right) \notag \\
&=&\left( \alpha +\beta \right) p^{0}p^{0\prime }|00\rangle +\left( \alpha
p^{0}\left( q_{0}^{\prime }+\rho p^{0\prime }\right) +\beta
p^{0}q_{0}^{\prime }\right) 01\rangle \notag \\
&&+\left( \alpha q_{0}p^{0\prime }+\beta \left( q_{0}+\rho p^{0}\right)
p^{0\prime }\right) |10\rangle +\left( \alpha q_{0}\left( q_{0}^{\prime
}+\rho p^{0\prime }\right) +\beta \left( q_{0}+\rho p^{0}\right)
q_{0}^{\prime }\right) |11\rangle.
\notag \\
\label{30-nov}
\end{eqnarray}%

In the limit of {indistinguishability} of the KK EBH states, one can
generate different states of the composed KK two-fold Hilbert space by a
suitable choice of the complex parameters $\alpha $ and $\beta $; indeed, in~such a limit, Equation (\ref{30-nov}) simplifies down to:%

\begin{eqnarray}
\left. \hat{\mathcal{E}}_{PQ}\right\vert _{c_{i}=0}\left( |\mathcal{Q}%
_{KK}\rangle \otimes |\mathcal{Q}_{KK}\rangle \right) &=&\left( \alpha
+\beta \right) \left( p^{0}\right) ^{2}|00\rangle +\left( \left( \alpha
+\beta \right) p^{0}q_{0}+\alpha \rho \left( p^{0}\right) ^{2}\right)
01\rangle \notag \\
&&+\left( \left( \alpha +\beta \right) p^{0}q_{0}+\beta \rho \left(
p^{0}\right) ^{2}\right) |10\rangle +\left( \alpha +\beta \right)
q_{0}\left( q_{0}+\rho p^{0}\right) |11\rangle. \label{30-nov-2}
\end{eqnarray}%

Thence, for instance, the conditions:%
\begin{equation}
\left\{
\begin{array}{l}
\left( \left( \alpha +\beta \right) p^{0}q_{0}+\alpha \rho \left(
p^{0}\right) ^{2}\right) =0, \\
\left( \left( \alpha +\beta \right) p^{0}q_{0}+\beta \rho \left(
p^{0}\right) ^{2}\right) =0%
\end{array}%
\right.
\end{equation}%
solved by:%
\begin{equation}
\alpha =\beta ,~~\rho =-2\frac{q_{0}}{p^{0}} \label{jazz}
\end{equation}%
yield a non-normalized Bell (GHZ) state:%
\begin{equation}
2\alpha \left( \left( p^{0}\right) ^{2}|00\rangle -q_{0}^{2}|11\rangle
\right).
\end{equation}%
on which the further normalization conditions:%
\begin{equation}
\left( p^{0}\right) ^{2}=\frac{1}{2\alpha \sqrt{2}}=-q_{0}^{2}
\end{equation}%
have a solution only when considering the {complexification} $\mathbf{%
R}_{\mathbb{C}}$ of the $G_{4}$-representation space of the EBH
electromagnetic charges. As observed, {e.g.,} in \cite{borsten0}, this
is a crucial step in order to implement a~quantum mechanical computation
(such as the ones exploited by the ``EBH quantum circuits''
and ``EBH quantum gates'' \cite{prudencio1}), since in QIT,
all parameters are generally complex and not real (as~instead, the electric
and magnetic charges of an EBH are). This clearly implies that the PQ
symplectic transformations themselves should be considered on a complex
ground field $\mathbb{C}$, since they act on the complex vector space $%
\mathbf{R}_{\mathbb{C}}$, on which $G_{4}(\mathbb{C})$ acts linearly, but
non-transitively. Thus, by choosing \footnote{
Note~that the ``$\pm $'' branches of $p^{0}$ and $q_{0}$ are
independent, but the ``$\pm $'' branch of $\rho $ depends on
their choice, consistently with Equation (\ref{jazz}).}:
\begin{equation}
\left\{
\begin{array}{l}
\alpha =\beta \\
p_{\pm }^{0}=\pm 2^{-3/4}\alpha ^{-1/2} \\
q_{0,\pm }=\pm i2^{-3/4}\alpha ^{-1/2} \\
\rho =\pm 2i%
\end{array}%
\right. \label{sols-1}
\end{equation}%
one obtains:%
\begin{equation}
\left. \hat{\mathcal{E}}_{PQ}\right\vert _{c_{i}=0}\left( |\mathcal{Q}%
_{KK}\rangle \otimes |\mathcal{Q}_{KK}\rangle \right) =\frac{1}{\sqrt{2}}%
\left( |00\rangle +|11\rangle \right) , \label{Bell}
\end{equation}%
namely, one can generate a normalized Bell (GHZ) state acting with the
entanglement PQ operator on a suitable, complexified ``large'' KK EBH state.

On the other hand, the conditions:%
\begin{equation}
\left\{
\begin{array}{l}
\left( \alpha +\beta \right) \left( p^{0}\right) ^{2}=0, \\
\left( \alpha +\beta \right) q_{0}\left( q_{0}+\rho p^{0}\right) =0%
\end{array}%
\right.
\end{equation}%
solved by:%
\begin{equation}
\alpha =-\beta \label{jazz-2}
\end{equation}%
yield a non-normalized W-state:%
\begin{equation}
-\alpha \rho \left( p^{0}\right) ^{2}\left( |10\rangle -|01\rangle \right) ,
\end{equation}%
which cannot be normalized, even on $\mathbb{C}$: necessarily, a phase $%
e^{i\pi }$ is introduced between the pure states $|10\rangle $ and $%
|01\rangle $. Indeed, by choosing:%
\begin{equation*}
\alpha =-\frac{1}{\sqrt{2}\rho \left( p^{0}\right) ^{2}}=-\beta ,
\end{equation*}%
one obtains:%
\begin{equation}
\left. \hat{\mathcal{E}}_{PQ}\right\vert _{c_{i}=0}\left( |\mathcal{Q}%
_{KK}\rangle \otimes |\mathcal{Q}_{KK}\rangle \right) =\frac{1}{\sqrt{2}}%
\left( |10\rangle -|01\rangle \right) , \label{W-phase}
\end{equation}%
namely, one can generate a normalized W-state {with }$e^{i\pi }$%
{-interference} acting with the entanglement PQ operator on a
suitable (not necessarily complexified!
) ``large'' KK EBH
state. The presence of the interference between the pure states $|10\rangle $
and $|01\rangle $ in Equation (\ref{W-phase}) seems to be a feature of the
entanglement generation through PQ symplectic transformations, in the
indistinguishability limit/assumption.

However, we anticipate that when relaxing the indistinguishability
assumption ({i.e.}, when resorting to a multi-centered EBH picture),
much more freedom is introduced in the above entanglement generation
procedure, and more general (and less constrained) solutions are obtained
.

Clearly, the presence of other EBH charge configurations/states (such as the
ones considered in Sections \ref{Sec-3} and \ref{Sec-4}) allows for a wealth of
entangled EBH states. A thorough analysis of PQ entanglement generation on
EBH states will be carried in a forthcoming paper.

\section{\label{Conclusion}Conclusions}

Within the black-hole/qubit correspondence, in the present paper, we have
paved the way to a~consistent generation of entanglement in EBH systems, by
means of an operator implementation of the so-called Peccei--Quinn (PQ)
symplectic transformations.

Given an EBH state described by its supporting electromagnetic charge
configuration ({i.e.},~by~an~element, with a well-defined associated
rank, of the corresponding Freudenthal triple system), the action of the PQ
operator can result in various types of {orbit transmutations},
namely in various mechanisms of switching among different $U$-duality orbits.

In the indistinguishability assumption, the need for complexification of the
$U$-duality representation space $\mathbf{R}$ (resulting in a
complexification of the PQ transformations, as well) and the unavoidable
occurrence of phases among pure EBH states has been pointed out; however, we
anticipate that if one~resorts to the multi-centered EBH space-time picture
and relaxes the indistinguishability condition, {at~least} some of such
limits can be circumvented.

\section{Acknowledgements}

This work is partially under support of the projects Enxoval-PPPG-N03/2014-UFMA,
APCINTER-00273/14-FAPEMA, UNIVERSAL-01401/16-FAPEMA and UNIVERSAL-CNPq (Brazil). D.J.
 Cirilo-Lombardo is also grateful to Bogoliubov Laboratory of Theoretical Physics-JINR (Russia)
 and CONICET-(Argentina) for financial support.



\end{document}